\documentclass[conference]{IEEEtran}
\AtBeginDocument{%
  \providecommand\BibTeX{{%
    \normalfont B\kern-0.5em{\scshape i\kern-0.25em b}\kern-0.8em\TeX}}}

\usepackage{alltt}
\usepackage{listings}
\usepackage{graphicx}
\usepackage{url}
\usepackage{tabularx}
\usepackage[justification=centering]{caption}

\newcommand\blfootnote[1]{%
  \begingroup
  \renewcommand\thefootnote{}\footnote{#1}%
  \addtocounter{footnote}{-1}%
  \endgroup
}

\begin{document}

\title{Optimization and Amplification of Cache Side Channel Signals}

\author{\IEEEauthorblockN{David A. Kaplan}
\IEEEauthorblockA{Advanced Micro Devices, Inc.\\Email: david.kaplan@amd.com}
}

\maketitle
\thispagestyle{plain}
\pagestyle{plain}
\blfootnote{ \textcopyright 2023 Advanced Micro Devices, Inc.  All rights reserved.
AMD, the AMD Arrow logo, and combinations thereof are trademarks of Advanced Micro Devices, Inc.
Other product names used in this publication are for identification purposes only and may be trademarks of their respective companies.}

\begin{abstract}

In cache-based side channel attacks, an attacker infers information about the victim based on the presence, or lack thereof, of one or more cachelines.  Determining a cacheline's presence, which we refer to as "reading the signal", typically requires testing the access time of the line using a suitably high precision timer.  In this paper we introduce novel gadgets which leverage CPU speculation to enable modification of these signals, before they are read, for a variety of purposes.  First, these gadgets enable an attacker to optimize cache-based side channel attacks by evaluating arbitrary logic functions on cacheline signals prior to their measurement.  Second, we demonstrate amplification techniques that enable an attacker to read a signal even if no high precision timer is available.  Combined, these techniques can be used to improve existing side channel attacks even if timer access is limited.  We evaluate the effectiveness of these techniques on a modern x86 CPU and demonstrate that when properly tuned, cache side channel signals can be reliably modified with near 100\% accuracy and are able to be read with a timer as coarse as 100$ms$ or more.

\end{abstract}

\section{Introduction}
The CPU cache has been widely used for side channels attacks in a variety of research.  Side channel attacks have previously been demonstrated in the recovery of sensitive data (including AES\cite{aesattack,aes2,aes3,aes4} and RSA\cite{rsa_cache_timing} keys) in vulnerable cryptographic libraries.  In 2018, Google Project Zero\cite{gpz} and independent researchers\cite{spectre} demonstrated a variety of speculation-based side channel attacks that leveraged the CPU cache to leak information from one program or protection domain to another.  As the cache is a shared resource, side channel attacks  have been demonstrated at the L1, L2 and L3\cite{llccache,llc2} levels on modern CPUs.

A variety of techniques for cache-based side channel attacks have been studied, including PRIME+PROBE\cite{primeprobe}, FLUSH+RELOAD\cite{flushreload}, and others\cite{takeaway}.  Regardless of the specific technique used, the attack typically works in 3 stages.  First, the attacker puts the system into a desired state, perhaps by loading or evicting certain cache lines.  Second, the victim is invoked resulting in some manipulation of the cache state. Finally, the attacker attempts to detect the manipulation of the cache state, thereby inferring information about the victim. 

This paper focuses solely on the final stage of these attacks, which we call signal recovery.  In the most basic FLUSH+RELOAD attack, a single cacheline may or may not be present, which we define as a signal of 1 or 0 respectively.  The attacker typically reads the signal value by timing the access to this cacheline using the x86 RDTSC instruction, the \verb|performance.now| JavaScript function, or another suitably high precision timer.  The attack is often repeated many times in order to read more information and eliminate potential noise.

There are a variety of practical challenges an attacker faces when attempting to do signal recovery on modern systems.  In response to Spectre, most web browsers have taken steps to dramatically reduce the precision of \verb|performance.now| and eliminate user-created high precision timers created via SharedArrayBuffer.  Microsoft\cite{msftblog} and Mozilla\cite{mozillaadvisory} initially reduced the resolution of \verb|performance.now| to 20$us$ with variable jitter of an additional 20$us$.  WebKit\cite{webkitblog} further reduced the resolution of \verb|performance.now| to 1$ms$.  As the time difference between a single cache hit and a miss is typically less than 100$ns$, this creates significant signal recovery challenges for any type of cache-based side channel attack.

Even if high precision timers are available to the attacker, signal recovery optimization may still be desired.  Mitigations have been proposed\cite{cache1,timewarp} to look for abnormally high numbers of calls to high precision timers (like RDTSC) to try to detect an in-progress side channel attack.  In this paper we demonstrate ways to manipulate cache side channel signals \textit{without reading them} thereby reducing the number of timing checks eventually needed.  Besides avoiding detection, this may also lead to optimized attack implementations.

To our knowledge, this is the first paper to focus exclusively on this signal recovery aspect of cache side channels.  Our specific contributions include:
\begin{itemize}
    \item Introducing a generic primitive that operates on cache side channel signals without requiring a time source
    \item Introducing gadgets which use this generic primitive to perform logical operations on cache side channel signals on x86 platforms
    \item Demonstrating signal amplification techniques which allows for cache signal recovery with a timer granularity of over 100$ms$
\end{itemize}

The rest of the paper is organized as follows.  Section \ref{background} discusses the underlying hardware and related work in this space.  Section \ref{primitive} introduces the generic primitive we use for signal modification with specific gadgets discussed in Section \ref{gadgets}.  Section \ref{amplification} and \ref{usecases} discuss the theory behind signal amplification and applications of signal modification.  And Section \ref{eval} demonstrates the gadgets and algorithms on a modern x86 system.  At the end of the paper, we briefly discuss potential mitigations and opportunities for further research.

~\\
\textbf{Note:} During the development of this paper, we became aware of similar research being conducted independently by another research team \cite{GatesOfTime}.  We wish to acknowledge this concurrent work, and thank the authors for their coordination and discussions around this topic.  While the publication of the work by both teams were coordinated, the content and results presented in this paper represent the sole work of the listed author.

\section{Background And Related Work} \label{background}
\subsection{Cache Side Channels}
Whenever a hardware resource is shared among two or more entities, there is a possibility of side and covert channels.  The CPU cache hierarchy is a resource commonly used in side channel attacks since in modern CPU designs the L1/L2 caches are typically shared between sibling hardware threads\cite{hyperthreading} in an SMT architecture, and the L3 cache is typically shared between multiple physical processors.  Additionally, software programs scheduled on the same physical CPU effectively share the same cache hierarchy.  As memory accesses that hit the cache execute faster than ones who do not, malicious programs can use the cache to communicate or spy on other programs, generally referred to as covert and side channel attacks, respectively.

A variety of research \cite{cachesurvey} exists around the use of cache-based side channels.  The general theme in a side channel attack is that during the execution of the victim program, it perturbs the cache in a way that is noticeable by the attacker.  If the way the victim perturbs the cache is directly related to a secret, such as a cryptographic key, the attacker may be able to determine the value of that secret.

Side channel attacks may generally be classified as either architectural or speculative.  Architectural side channel attacks involve an attacker determining secrets from the architectural (non-speculative) flow of software execution.  Architectural side channel attacks historically have been addressed by software updates, such as by removing secret-dependent memory access patterns\cite{memory_access_patterns}.

Speculative side channels, such as Spectre\cite{spectre}, Meltdown\cite{Lipp2018meltdown},  Foreshadow\cite{foreshadowNG}, and MDS\cite{ridl, fallout} rely on the speculative flow of execution in CPU hardware.  These attacks leverage features like branch prediction and out-of-order execution and represent a newer class of side channels.  In response to these, a variety of mitigations have been deployed including software and hardware updates\cite{intel_spectre_wp,amd_spectre_wp}.

In both architectural and speculative cache-based side channels, the attacker determines the victim's secret through the timing of memory accesses.  Whether a particular address is or is not cached tells the attacker something about the victim.  Because modern CPUs operate at very high frequencies, a very high precision timer is necessary for determining whether a particular address is cached or not.  A common technique is to use the x86 Time Stamp Counter (TSC) which counts CPU cycles.  By using instructions like \verb|RDTSC|\cite{apmvol3} around a memory access, an attacker can get a very accurate timing of how long that access took to execute.

As discussed earlier, instructions like \verb|RDTSC| may not always be available to an attacker, or their use may need to be minimized to avoid detection.  In this paper, we demonstrate how CPU speculation can be used to achieve these goals with new signal modification and recovery primitives.

\subsection{CPU Speculation}
Speculation is a common technique used in high performance CPU design and involves the CPU starting to evaluate instructions before it is determined whether those instructions are architecturally part of the program flow.  A variety of speculation techniques are used in modern microprocessors, including branch prediction\cite{1981bpstudy}.  In branch prediction, the CPU hardware guesses how a branch will be evaluated and begins speculatively executing the instructions along the chosen path.  If the hardware later determines the guess was incorrect, the architectural effects of the speculative execution are reversed and in-progress instructions are flushed, allowing the correct path to begin execution.  However micro-architectural effects of speculation can sometimes remain, including cachelines brought in as a result of speculation.

As modern processors employ out-of-order execution techniques, instructions including branches are evaluated independent of program order.  As long as the data dependencies may be satisfied, the branch will be evaluated and potentially redirect the execution of the pipeline.  If the data dependencies cannot be immediately satisfied, such as a branch dependent on a load from DRAM, the execution of the branch waits until the required data is available.  This behavior is key to the work presented here, as the length of time it takes to evaluate a branch is directly dependent on the time to load its dependent data.

\subsection{Related Work}
While much existing research\cite{cachesurvey} exists on CPU side channels and their applications, there appears to be very little peer-reviewed work in the general area of signal modification or amplification.  That being said, in response to the Spectre browser mitigations\cite{msftblog,mozillaadvisory,webkitblog} deployed in early 2018, there have been various workarounds proposed for the lack of high precision timers.  These typically include constructing artificial high-precision timers using available resources\cite{timers}, or calling the victim code multiple times to cause several cache lines to be affected and hitting specific timing windows\cite{overcomingmitigations}.  Another technique involves using a self-reinforcing loop \cite{spectrecascade} and CPU speculation to amplify the original signal, while also repeatedly executing the Spectre gadget.

Unlike these techniques, the work presented here does not rely on artificial high-performance timers, nor invoking the victim code multiple times.  While invoking the sensitive code repeatedly may be practical for Spectre-based attacks (such as an out-of-bounds array load), in other attacks the attacker may need to hit a very small timing window to observe the victim's behavior.  For instance, the MDS attack may need to execute around the exact time when the secret is being used.  In these cases, it is likely not practical to attempt to hit this timing window multiple times and instead the attacker must be able to determine as much information as possible from a single invocation.  This paper demonstrates methods for amplifying signals to the point that high precision timers are not needed, as well as methods for modifying signals prior to measurement to reduce the number of timing checks needed.  These techniques can apply to most types of cache-based side or covert channel attack.

\section{Definitions} \label{definitions}
Throughout this paper, we will refer to cachelines by their virtual address.  As this paper focuses on manipulating the state of cachelines, we define $\phi(A)$ to return TRUE (1) if the cacheline at address $A$ is present in any level of the CPU cache hierarchy and FALSE (0) otherwise.  More specifically, given a microarchitecture specific threshold value T, $\phi(A)=1$ if the time required to access $A$ is less than T, and $\phi(A)=0$ otherwise.  

For purposes of signal recovery, the value of the data in the cache typically does not matter and is controlled by the attacker.  For instance, in a FLUSH+RELOAD attack, the attacker is interested to know if a certain cache line is present or not, but the value of the data is not relevant.  Therefore, in the examples presented in this paper we universally assume that all cachelines signals used contain the value 0.  That said all code examples shown here could be modified as needed if this assumption was not possible.

All assembly code in this paper is written in x86 GNU Assembler syntax\cite{gnu_syntax}.  In GNU Assembler syntax, the destination is always on the right (e.g., $add\  src, dest$).

\section{Basic Primitive} \label{primitive}
\subsection{Theory}
In order to manipulate cache signals without reading them, we take advantage of CPU speculation in a controlled manner.  The key insight is that we can make the length of speculation completely dependent on the presence (or lack thereof) of one or more cachelines.  To demonstrate how this works, let us first examine how to build a simpler inverter.

Assume that initially cacheline $B$ is not present in the cache and the line pointed to by $A$ may or may not initially be present.  As noted above, memory at $A$ is assumed to contain the value 0.  The most basic form of the signal manipulation primitive is the inverter shown below:

\begin{alltt}
    if (likely(*A != 0)) \{
        Access cacheline B
    \}
\end{alltt}

The idea is that the CPU is first tricked into mispredicting the branch, resulting in cacheline B being accessed during speculation and potentially brought into the cache.  Eventually the CPU will discover that it mispredicted the branch and stop the incorrect speculation.  How many cycles it takes for the CPU to discover the misprediction is dependent on whether cacheline A was initially present or not.  If cacheline A was present, the branch evaluates quickly and it is unlikely cacheline B will be fetched.  But if cacheline A was not present, the branch evaluates slowly resulting in the access to cacheline B.

\subsection{Branch Misprediction}
To realize this primitive in practice, there are a few challenges to resolve.  First is how to ensure the CPU mispredicts the branch to achieve the desired speculation.  Although this could be accomplished through sufficient priming of the conditional branch predictor, in this paper we take advantage of a common and more reliable behavior of most CPUs whereby $ret$ instructions are typically predicted to return to the instruction following the previous $call$.   This is the same insight that forms the basis for the $retpoline$ \cite{retpoline} Spectre mitigation.  The code sequence used is shown in Figure \ref{fig:asm_forced_spec}.

\begin{figure}
    \centering
\begin{lstlisting}[language={[x86masm]Assembler},frame=lines]

    1: call 3f
    2: #Speculative instructions go here
       lfence
    3: mov $4f, (%rsp)
       ret
    4: nop
\end{lstlisting}

   \caption{Forced Speculation Gadget}
    \label{fig:asm_forced_spec}
\end{figure}

In the architectural flow of this code, the return address on top of the stack is  overwritten to point to label 4 so when the $ret$ executes, the processor will jump to the $nop$ instruction.  However, the branch prediction logic will not be aware of this and predict that the $ret$ will return back to label 2.  The instructions at label 2 will therefore be executed speculatively, and this occurs every time this code is executed without requiring any kind of priming. 

\subsection{Inverter}
 \begin{figure}
     \centering
\begin{lstlisting}[language={[x86masm]Assembler},frame=lines]
    1: call 3f
    2: xor %rax, %rax
       .rept 5 # Delay ops
       mov (%rsp, %rax), %rax
       and $0, %rax
       .endr
       mov (%rdi, %rax), %r11
       lfence
    3: mov $4f, (%rsp)
       mov (%rsi), %r11
       add %r11, (%rsp)
       ret
    4: nop
\end{lstlisting}

 \caption{Inverter}
     \label{fig:asm_inverter}
 \end{figure}

\begin{figure*}[t]
    \centering
    \frame{\includegraphics{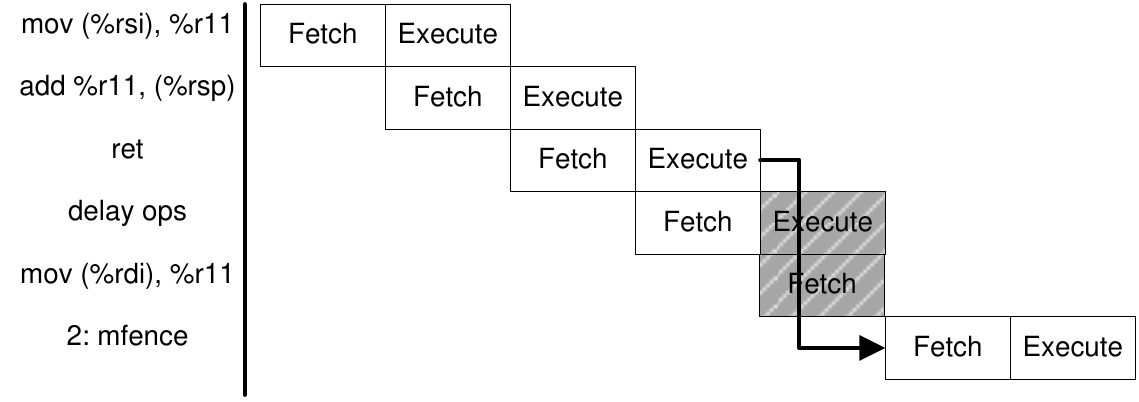}}
    \caption{Pipeline (Cacheline was present)}
    \label{fig:pipeline_fast}
\end{figure*}
\begin{figure*}[t]
    \centering
    \frame{\includegraphics{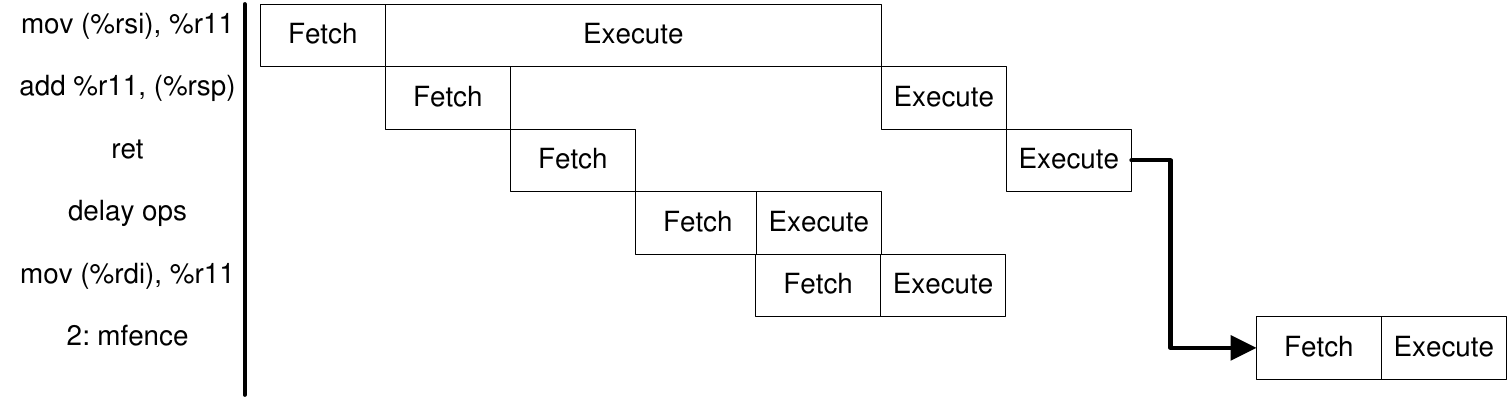}}
    \caption{Pipeline (Cacheline was not present)}
    \label{fig:pipeline_slow}
\end{figure*}

Using the terminology introduced earlier, the function of the signal inverter is that given an input cacheline A and output cacheline B, set $\phi(B) = \neg\phi(A)$.  In the code shown in Figure \ref{fig:asm_inverter}, the address of A is assumed to be in RSI while the address of B is in RDI.  Note that cacheline B is assumed to initially not be present ($\phi(B)=0$).

There are two important elements of the Inverter code.  The first is that a data dependency is created between cacheline A (RSI) and the top of the stack.  The addition of the value of cacheline A (assumed to be 0) does not change the value of the new return address being written, but the CPU does not know this.  It is not able to evaluate the $ret$ until this dependency is satisfied.

The second aspect is the delay loop shown at label 2.  This is needed because even if the processor is quickly able to evaluate the $ret$, some ops may have executed along the incorrect speculative path.  This delay loop performs dependent memory operations in order to give the CPU time to redirect itself if the input cacheline happened to be present.

The simplified pipeline diagrams in Figure \ref{fig:pipeline_fast} and \ref{fig:pipeline_slow} demonstrate this behavior.  In Figure \ref{fig:pipeline_fast}, the execution of the load from RSI happens quickly allowing the $ret$ to flush the pipeline before the load of RDI begins to execute.  However in Figure \ref{fig:pipeline_slow}, the load from RSI happens slowly, and the RDI load is able to execute. 

The result is an inversion of $\phi(A)$.  If $\phi(A)=0$ then the $ret$ is slow and cacheline B is accessed (so $\phi(B)=1$).  But if $\phi(A)=1$ then the $ret$ is fast and cacheline B is not accessed.

\section{Logic Gadgets} \label{gadgets}
This section demonstrates several examples for manipulating the state of cachelines to form standard boolean logic gates, building upon the basic speculation primitive.
\subsection{Replicator} \label{replicator}
The replicator is similar to the inverter but aims to bring in more than one cacheline based on the input.  For instance, with three output lines it sets $\phi(B_1)=\phi(B_2)=\phi(B_3)=\neg\phi(A)$.  When accessing the output lines, these instructions are each made to be dependent on the delay loop, but independent of each other.  This allows their lines to be fetched in parallel.  The code in Figure \ref{fig:asm_replicator} implements the replicator with an output of three lines.  In this example, the target cachelines are pointed to by R8, R9, and R10.

 \begin{figure}
     \centering
\begin{lstlisting}[language={[x86masm]Assembler},frame=lines]
    1: call 3f
    2: <delay ops>
       mov (%r8, %rax), %r11
       mov (%r9, %rax), %r11
       mov (%r10, %rax), %r11
       lfence
    3: mov $4f, (%rsp)
       mov (%rsi), %r11
       add %r11, (%rsp)
       ret
    4: nop
\end{lstlisting}

 \caption{Replicator}
     \label{fig:asm_replicator}
 \end{figure}

\subsection{NAND Gadget}
In the above examples, only a single cacheline served as the input.  Next we modify the primitive to look at multiple input cachelines, enabling boolean logic.  For instance, the code in Figure \ref{fig:asm_nand} sets $\phi(C)=\neg(\phi(A) \wedge \phi(B))$.  Cachelines A, B, and C are assumed to be in registers RSI, RDX, and RDI respectively.

 \begin{figure}
     \centering
\begin{lstlisting}[language={[x86masm]Assembler},frame=lines]
    1: call 3f
    2: <delay ops>
       mov (%rdi, %rax), %r11
       lfence
    3: mov $4f, (%rsp)
       mov (%rsi), %r11
       add (%rdx), %r11
       add %r11, (%rsp)
       ret
    4: nop
\end{lstlisting}
 \caption{2-input NAND}
     \label{fig:asm_nand}
 \end{figure}

In this gadget, the CPU is only able to send the redirect and flush the pipeline if both input cachelines are present.  If either one of them is not present, speculation should continue far enough to load the output cacheline.

\subsection{NOR Gadget}
As another example, a NOR gate which sets $\phi(C)=\neg(\phi(A) \vee \phi(B))$ can be built using a primitive where either cacheline being present causes the CPU to quickly end the incorrect speculation.
In the NOR gate, there are 2 instances of forced incorrect speculation.  The first one ($call\ 4f$) will end quickly if $\phi(A)=1$ while the second ($call\ 5f$) will end quickly if $\phi(B)=1$.  The output cacheline C will only be accessed therefore if both $\phi(A)=\phi(B)=0$.   This gadget is shown in Figure \ref{fig:asm_nor}.  Cachelines A, B, and C are assumed to be in registers RSI, RDX, and RDI respectively.

\begin{figure}
    \centering
\begin{lstlisting}[language={[x86masm]Assembler},frame=lines]
    1: call 4f
    2: call 5f
    3: <delay ops>
       mov (%rdi, %rax), %r11
       lfence
    4: mov $6f, (%rsp)
       mov (%rsi), %r11
       add %r11, (%rsp)
       ret
    5: mov $6f, (%rsp)
       mov (%rdx), %r11
       add %r11, (%rsp)
       ret
    6: nop
\end{lstlisting}
    \caption{2-input NOR}
    \label{fig:asm_nor}
\end{figure}

\subsection{Arbitrary Logic}
As both NAND and NOR are universal logic functions, it is possible to implement arbitrary logic functions on the state of cachelines using either of these gadgets.  The gadgets shown here may also be trivially extended to provide additional inputs (e.g., a 4-way NAND) or additional outputs.  There are limits on how many inputs and outputs may be used in a single gadget in a given microarchitecture and this is discussed more later in section \ref{amplification}.

Note that it is possible to directly build a 2-input XOR gadget using the primitive with three branch mispredictions.  It functions by accessing the output cacheline if and only if exactly one of the input cachelines can be accessed quickly.  As will be discussed in the evaluation section however, this gadget did not perform as reliably as the other gadgets presented here.

\section{Signal Amplification} \label{amplification}
Signal amplification refers to the problem of determining the state of a cacheline without access to a high precision timer.  In this case, we assume there is an initial cacheline A in an unknown state.  Our goal is to manipulate the cache such that when we eventually perform a timing measurement, the timing difference between the case where $\phi(A)$ was 0 vs 1 is as high as possible.  We refer to this timing difference as the signal strength.

\subsection{Amplifier Theory}
The basic idea of amplification is to make the state of multiple cachelines dependent on the single input.  We can then time the access to those multiple cachelines and observe a larger timing difference than the access to the single cacheline.

Let us assume we have an initial cacheline A and two additional cachelines B and C to work with.  Using the replicator shown in Section \ref{replicator}, we can set the state of cachelines B and C to be the inverse of A.  To create a noticeable timing delta though, we must be careful in how cachelines are accessed.

In the replicator gadget, cachelines B and C should be accessed independently (meaning the access to cacheline C is not dependent on the result of the access to cacheline B).  This ensures the processor is able to perform both cache accesses in parallel.  However when the timing measurement is performed, cachelines B and C should be accessed in a dependent manner.  That is, the access to cacheline C $should$ be dependent on the access to cacheline B.  This ensures that the processor is not able to perform these accesses in parallel, thus exacerbating the timing difference between cache hits and cache misses.

\subsection{Single-Stage Amplifier}
The single stage amplifier using a single round of speculation to bring in as many cachelines as possible based on the input.  The amplification primitive is parameterized by two values, $deplen$ and $accesslen$.  $deplen$ is defined as the length of the delay loop, while $accesslen$ is defined as the number of output cachelines potentially accessed.  In x86 code, this primitive is implemented as shown in Figure \ref{fig:asm_amp}.

\begin{figure}
    \centering
\begin{lstlisting}[language={[x86masm]Assembler},frame=lines]
    .macro amp deplen accesslen
    1: call 3f
    2: xor %rax, %rax
       .rept \deplen
       add (%rsp, %rax), %rax
       and $0, %rax
       .endr
       .rept \accesslen
       mov (%rdi, %rax), %r11
       add $STRIDE, %rdi
       .endr
       lfence
    3: mov $4f, (%rsp)
       mov (%rsi), %r11
       add %r11, (%rsp)
       ret
    4: nop
    .endm
\end{lstlisting}
    \caption{Single Stage Amplifier}
    \label{fig:asm_amp}
\end{figure}

The optimal single stage amplifier will use the smallest practical $deplen$ value with the highest practical $accesslen$ value.  The $deplen$ should be as small as possible while still ensuring that if the input cacheline (in RSI) is present, none of the cachelines starting at RDI are accessed.  The $accesslen$ value should be as large as a particular microarchitecture can support.  As discussed in Section \ref{eval}, an $accesslen$ over 20 can be possible.

In the example in Figure \ref{fig:asm_amp}, the output cachelines are separated by \verb|STRIDE| bytes.  The optimal value of this varies by microarchitecture and is described more in Section \ref{eval}.

\subsection{Multi-Stage Amplifier}
While a single invocation of the single stage amplifier can affect only a certain number of output cachelines, these amplifiers could be chained together to affect more cachelines to create a multi-stage amplifier.  This would theoretically result in the ability to create arbitrarily large signal strengths.

Unfortunately this technique does not work well in real systems because cache sizes are limited.  And due to other system activity, it is likely impossible to manipulate the entire state of the cache on a system without interference.  In practice, multi-stage amplifiers typically only work with a small number of stages, are subject to more unwanted noise, and generally do not yield signal strengths large enough to measure with extremely coarse timers.

\subsection{Self-reinforcing Amplifier}
\begin{figure}[t]
    \centering
    \includegraphics{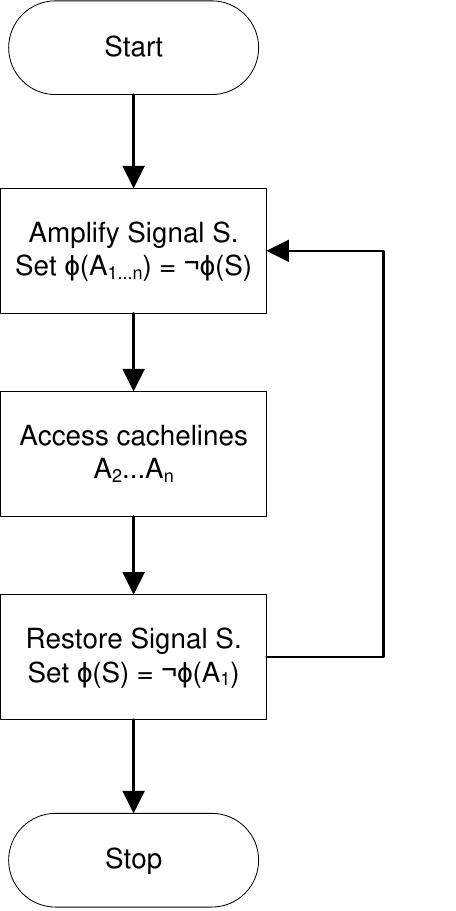}
    \caption{Self-reinforcing Amplifier}
    \label{fig:self_reinforce}
\end{figure}

A more scalable technique to signal amplification is through a self-reinforcing amplifier.  The idea of the self-reinforcing amplifier is to amplify the signal using the single stage amplifier, access all but one of the output lines, and finally use the remaining line to restore the original signal as shown in Figure \ref{fig:self_reinforce}.  As discussed earlier, during the access stage a dependency on the output cachelines is used to force the timing of the entire loop to become dependent on the original signal.  This loop can be repeated many times to create a large signal strength.

\section{Applications} \label{usecases}
Modifying cache signals prior to reading them can be useful for reducing the number of timer calls needed.  This is especially valuable is a high precision timer is either not available or if access to it is being monitored.

\subsection{FLUSH+RELOAD Binary Search} \label{binary_search}

\begin{figure}[t]
    \centering
    \frame{\includegraphics{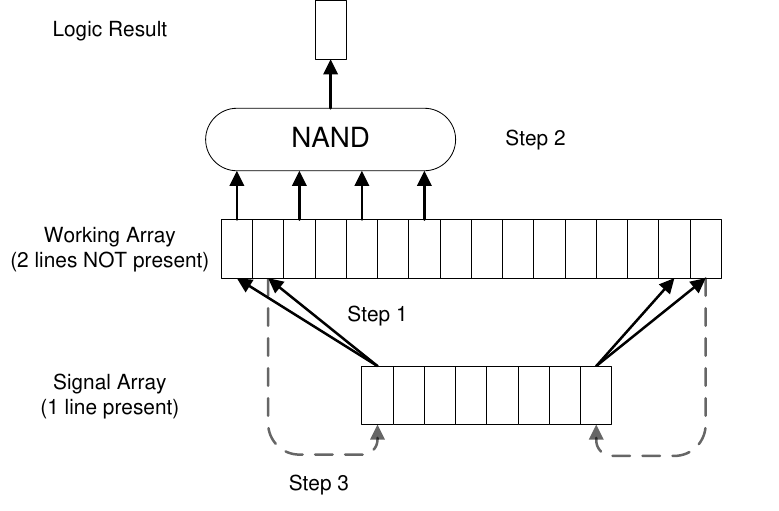}}
    \caption{Binary Search}
    \label{fig:binsearch}
\end{figure}

One use case for modifying signals prior to recovery is an efficient FLUSH+RELOAD binary search.  The premise is that the attacker is doing a FLUSH+RELOAD attack and has a set of N cachelines, exactly one of which will be brought in by the victim depending on the result of the attack.  Each of the cachelines represents a possible output value and each value is equally likely to occur.  A naive approach would be to time the access to each of the lines, looking for the one that is fast.  However we will show that the attacker can in fact accomplish this with only $log_2(N)$ timed accesses.

An example with N=16 is shown in Figure \ref{fig:binsearch}.  This algorithm will make use of three types of gadgets.  Specifically we will use an inverter, a 2x replicator, and a variety of NAND gadgets (8, 4, and 2-input versions).  Additionally we will use several memory arrays.  The first array, $S_0...S_{15}$ is the "signal" and we assume that exactly one of these cachelines is initially present.  The second array W is defined as twice the size of S, from $W_0...W_{31}$.  The last array is just a single cacheline in size, called R.

On each loop iteration, first the D and R arrays are flushed from the cache.  Then the array S is duplicated such that $\phi(W_{2*i}))=\phi(W_{2*i+1}))=\neg\phi(S_i)$, shown as Step 1.  Next a test is run to determine which half of the array must be searched.  To test indexes x through y we compute a NAND such that $\phi(R)=\neg(\phi(W_{2*x}) \wedge \phi(W_{2*(x+1)}...\wedge \phi(W_{2*y}))$, shown as Step 2.  Finally, we copy the unused lines in W back to S using the inverter such that $\phi(S_i)=\neg\phi(W_{2*i+1})$ as shown in Step 3.  This step is necessary because each primitive used is destructive to the input.  The replicator helps work around this by keeping an (inverted) copy of the original signal array in odd entries of W.

After each loop, we test if line R is present in the cache or not.  If R was present it means that the signal cacheline was in the range $x...y$ and we narrow our search range to those indexes.  If R was not present, the signal was in the other half.

As with a standard binary search, this algorithm is able to determine which of N cachelines was present using only $log_2(N)$ timing checks.

\subsection{Cacheline Counter} \label{cacheline_counter}
Another use case that can be optimized is when an attacker has an array of cachelines which may or may not be present, and they want to know how many are present.  This might occur if the attacker is doing a FLUSH+RELOAD attack and the victim will bring in different numbers of lines based on what code it executed.

Using the logic gates described earlier, it is possible to build an adder where the state of a set of cachelines represent the counter value.  For example, the attacker may have an array of 64 cachelines, in which case a 6-bit adder is needed.  The state of this adder is represented by six cachelines, each corresponding to a bit.  The attacker then takes each of the set of input cachelines and performs a binary addition between that input cacheline and the 6-bit adder.  

\begin{figure}[t]
    \centering
    \frame{\includegraphics[width=\columnwidth]{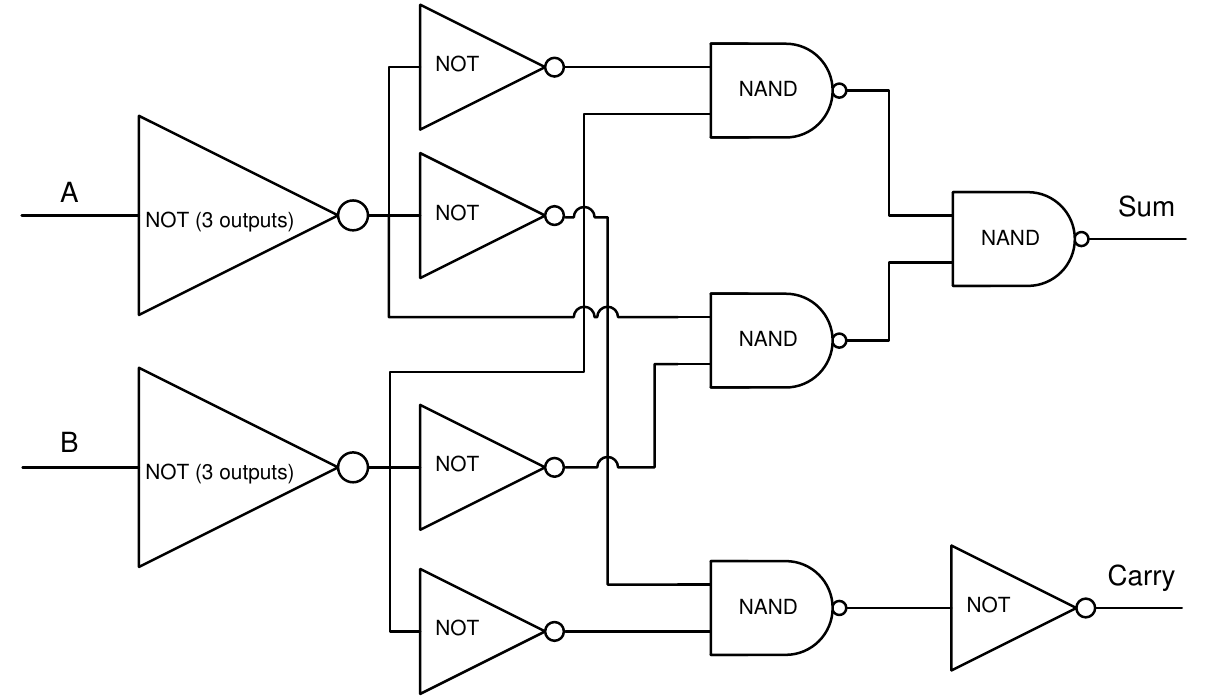}}
    \caption{Half Adder for Cacheline Signals}
    \label{fig:half_adder}
\end{figure}

The binary addition is performed using the logic gadgets introduced earlier.  In particular, an inverter, 3x replicator, and 2-input NAND are used to build a half-adder as shown in in Figure \ref{fig:half_adder}.  Note that this logic is slightly different from a standard half-adder because of the limitations of using cache line state.  Each use of a particular cache line is destructive so copies of each input must be made at the beginning.  Additionally, as shown earlier it is easiest to build gates like NAND instead of XOR, so the half-adder is written exclusively using NAND and inverter gadgets.

After the state of all the input cachelines have been added together, the attacker can read the value of the counter by timing accesses to it.  In the example above, the attacker therefore only needs to perform six timing measurements to determine how many of the 64 cachelines were present.

\section{Evaluation} \label{eval}
An AMD Threadripper\texttrademark\  3970X running Ubuntu 20.04 and Linux\textregistered\ kernel 5.10.0-rc7 was used for testing.  This system has 32 CPU cores (64 threads with SMT) each with 32kB of L1 cache (8-way set associative), 512kB of L2 cache (8-way set associative), and 128MB of L3 (16-way set associative).  The cacheline size at all levels of the hierarchy is 64 bytes.  The L1 and L2 caches are private to each core on this CPU while the L3 cached is shared amongst groups of eight cores.  During testing, all code was pinned to a single CPU core.  Note that while an AMD processor was used for evaluation, the techniques described in this paper are generic and are likely to behave similarly on processors from other vendors as well.

In all tests, different cachelines were configured to be at least 4160 (4096+64) bytes apart.  Keeping cachelines at least 4096 bytes apart was necessary to avoid unwanted interaction from hardware prefetchers.  Offsetting cachelines by an additional 64 bytes ensures that different cachelines would reside at different L1 indexes to avoid contention as the L1 index uses bits 11:6 of the virtual address.
\subsection{Logic Gates}
A special testbench was created to test the reliability of the basic logic gate gadgets.  This testbench runs each each set of possible inputs to each logic gate for a total of 800k runs.  After each run, the output cachelines were checked to determine if the logic gate performed correctly.  The results are shown in Table \ref{tab:logic_gates}.  

The accuracy of the logic gates was generally very high.  The worst performing gate was the native XOR gate.  The native XOR gate struggled to consistently produce the correct result in the \{0, 0\} input case.  We believe this is because if one cacheline is returned significantly faster than the other for some reason, it effectively aliases to either the \{0, 1\} or \{1, 0\} case.  For this reason, the half adder logic used NOT and NAND gates instead as described in section \ref{cacheline_counter}.

\begin{table}[t]
    \centering
    \begin{tabular}{|c|c|c|c|}
    \hline
    Gate & Inputs & Correct Results & Accuracy \\
    \hline
    NOT & 1 & 799998 & 100\% \\
    NOR & 2 & 799998 & 100\% \\
    NAND & 2 & 799996 & 100\% \\
    NAND & 4 & 799996 & 100\% \\
    NAND & 8 & 799999 & 100\% \\
    NAND & 16 & 799995 & 100\% \\
    NAND & 32 & 799995 & 100\% \\
    NAND & 64 & 799941 & 99.99\% \\
    NAND & 128 & 799649 & 99.96\% \\
    XOR & 2 & 772869 & 96.61\% \\
    Half Adder & 2 & 799994 & 100\% \\
    \hline 
    
    \end{tabular}
    \caption{Logic Gate Results (800k runs)}
    \label{tab:logic_gates}
\end{table}
\subsection{Amplifiers}
To determine the optimal parameters for the amplifier, testing was performed to determine the how quickly the CPU could resolve speculation and how many different cachelines could be accessed during speculation.  On the AMD Threadripper CPU used, it was determined that a $deplen$ (delay) value of 5 and an $accesslen$ value of 23 worked well.  In other words, in the single stage amplifier, 23 output cachelines were set to the inverted state of the input cacheline.

Each of the 23 output cachelines were exactly \verb|STRIDE| bytes apart.  As mentioned earlier, this value was set to 4160 bytes to avoid L1 index contention as well as to avoid the hardware prefetcher accessing cachelines on its own.

\begin{figure}
    \centering
    \includegraphics[width=\columnwidth]{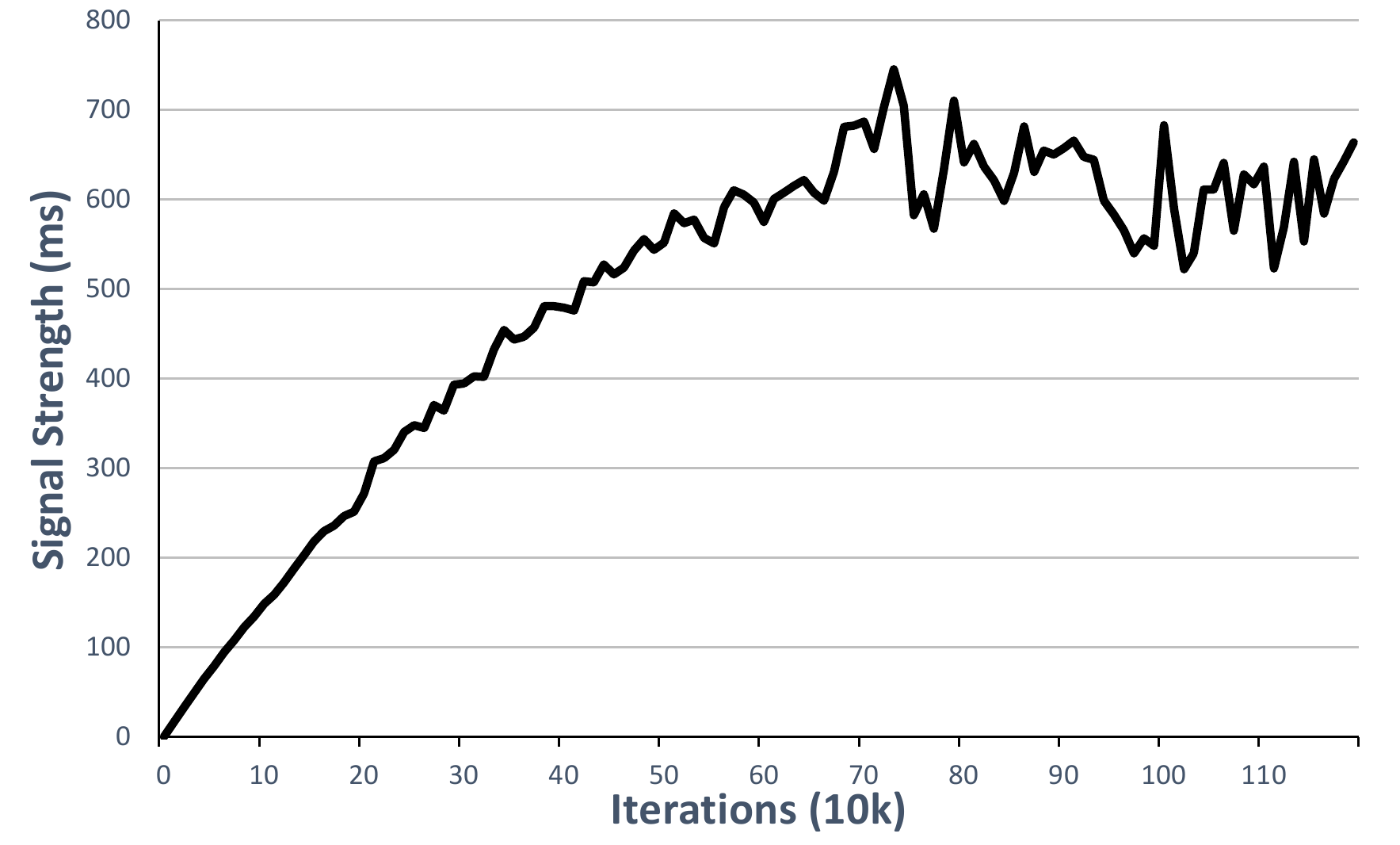}
    \caption{Amplifier Results (2000 runs each)}
    \label{fig:self_amp_results}
\end{figure}

The single stage amplifier was subsequently used to create a self-reinforcing amplifier.  The self-reinforcing amplifier was run for various numbers of iterations to determine the best possible signal strength as shown in Figure \ref{fig:self_amp_results}.

While the self-reinforcing amplifier could theoretically be run as many times as desired, the data indicates that the signal strength steadily rises until around 700k iterations after which additional iterations do not provide consistent benefits.  We believe this is because the gadgets used are subject to some risk of errors, such as due to system interrupts.  If the original signal gets corrupted, then the self-reinforcing amplifier will cease to function properly.  As the iteration count increases, the chance of signal corruption similarly grows.

Despite the risk of signal corruption, the results demonstrate that the self-reinforcing amplifier performs extremely well and is capable of amplifying signals above 600$ms$.  As a point of reference, on the tested platform a non-amplified signal had a strength of approximately 76$ns$ (as measured by the TSC).  This means that the amplification logic achieved a $8*10^6$ amplification on average.

\begin{figure}
    \centering
    \includegraphics[width=\columnwidth]{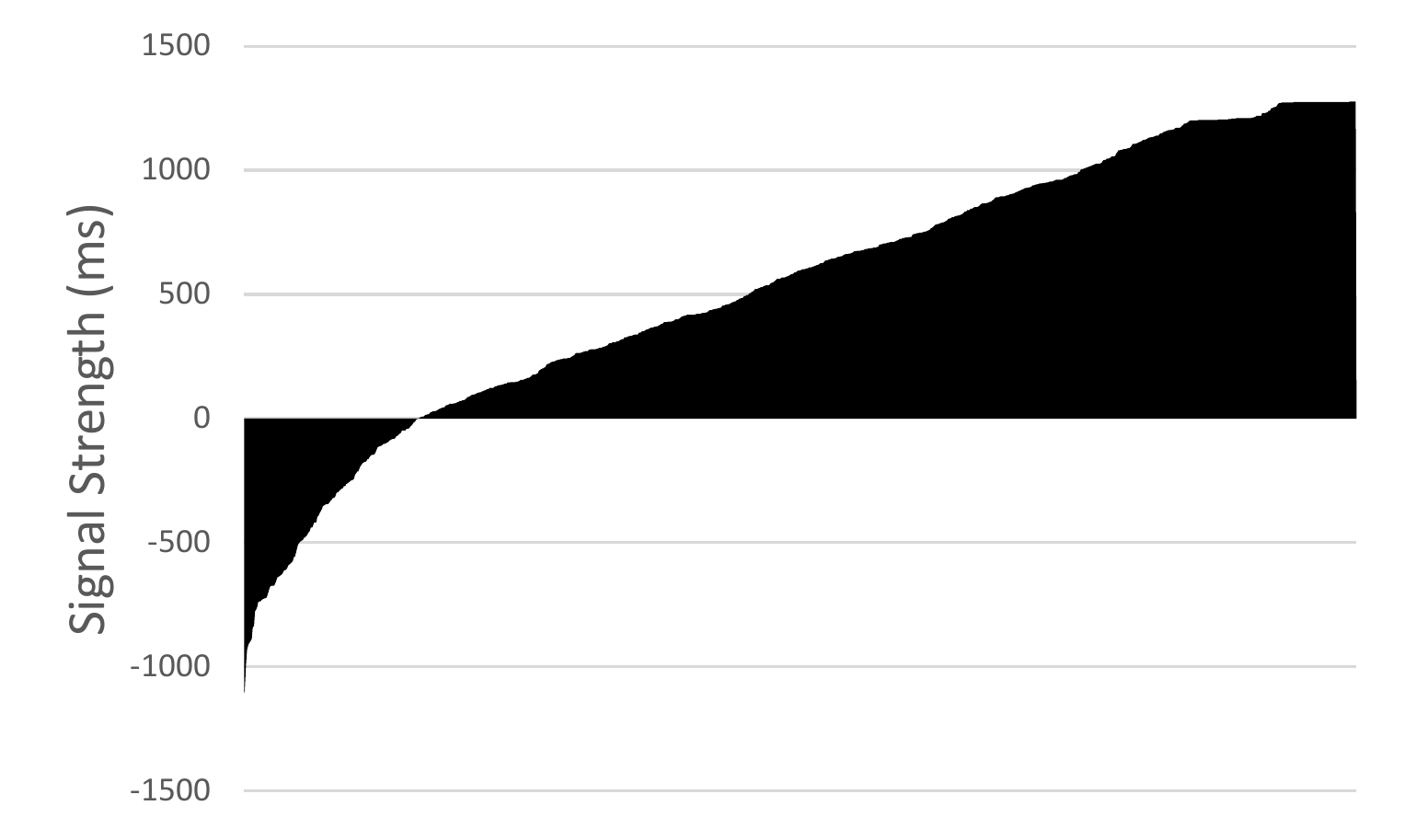}
    \caption{700k Iteration Amplifier Results (1000 runs)}
    \label{fig:700_results}
\end{figure}

\begin{figure}
    \centering
    \includegraphics[width=\columnwidth]{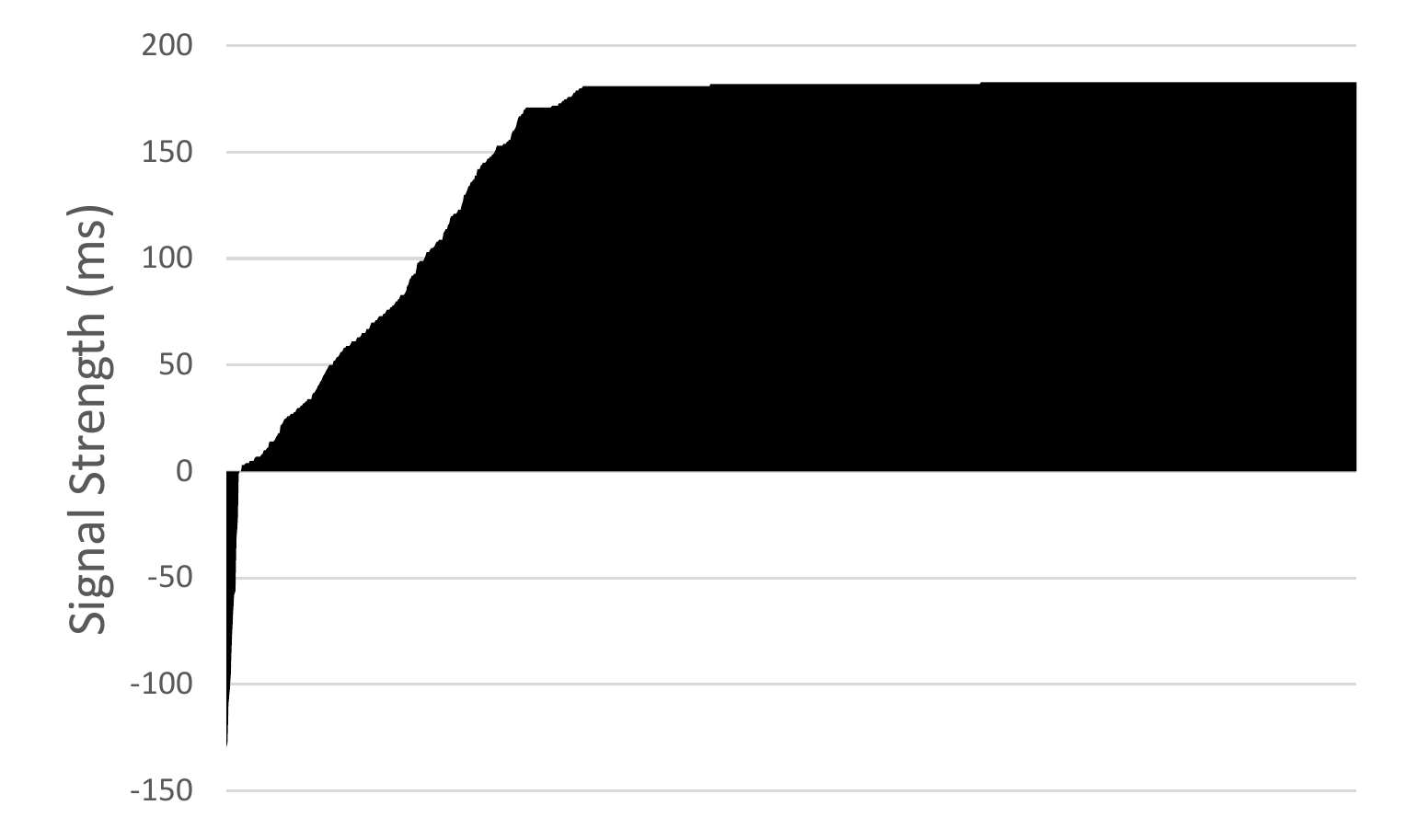}
    \caption{100k Iteration Amplifier Results (1000 runs)}
    \label{fig:100_results}
\end{figure}

While the average results of signal amplification are significant, the consistency of the algorithm was also evaluated.  Although average signal strength rises with higher iteration counts, there is also a wider range of results and higher risk of incorrect amplification.  Comparing the graphs of Figures \ref{fig:700_results} and \ref{fig:100_results} shows how despite 700k iterations yielding larger signal strengths, the results from 100k iterations were less subject to errors as indicated by fewer negative signal strengths.

The consistency of various iteration counts are shown in Table \ref{tab:amp_consistency}.  To determine the chance of correct signal recovery, the number of runs that demonstrated a signal above the given timer threshold were subtracted from the number of runs that demonstrated an incorrect signal below the same threshold.  For instance, in the 100k configuration, out of the 1000 runs tested, a signal strength above 10$ms$ was found in 964 runs.  But an incorrect signal below -$10ms$ was found in 11 runs.  In the remaining 25 runs the signal was less than $10ms$ and thus not able to be measured with a $10ms$ timer.  This result is reported in Table \ref{tab:amp_consistency} as a 95\% chance of correct signal recovery.

The chance of an indeterminate signal was defined as a result where the absolute signal strength was less than the granularity of the timer.  For example, in the 300k configuration with a 100$ms$ timer, there is an 18\% chance that the difference between the source cacheline being present or not present was measured as being less than 100$ms$.

The results show that the optimal choice of amplifier configuration is highly dependent on the available timer.  With a 10$ms$ timer the 100k configuration performs best among the ones tested as there is a 95\% chance of correct signal recovery.  On the other hand if only a 500$ms$ timer is available then the 700k amplifier is the best choice.  Although it only recovers the signal correctly 50\% of the time, the result is indeterminate another 40\% of the time.  The chance of an incorrect signal recovery is thus only 10\%.  If it is possible to conduct the attack multiple times, the chance of correct signal recovery will increase.

\begin{table}[t]
    \centering
    \begin{tabular}{|c|c|c|c|c|}
    \hline
    \textbf{Amplifier Configuration} & \textbf{100k} & \textbf{300k} & \textbf{500k} & \textbf{700k} \\
    \hline
    \multicolumn{5}{|c|}{Average Signal Strength ($ms$)} \\
    \hline
    1st Quartile & 155 & 112 & 175 & 155 \\
    Median & 182 & 293 & 484 & 596 \\
    3rd Quartile & 183 & 512 & 827 & 990 \\
    \hline
    \multicolumn{5}{|c|}{Chance of Correct Signal Recovery} \\
    \hline 
    With 500$ms$ timer & 0\% & 26\% & 45\% & 50\% \\
    With 100$ms$ timer & 82\% & 71\% & 72\% & 66\% \\
    With 10$ms$ timer &  95\% & 80\% & 75\% & 68\% \\
    \hline
    \multicolumn{5}{|c|}{Chance of Indeterminate Signal} \\
    \hline
    With 500$ms$ timer & 100\% & 74\% & 50\% & 40\% \\
    With 100$ms$ timer & 17\% & 18\% & 11\% & 8\% \\
    With 10$ms$ timer & 3\% & 2\% & 1\% & 1\% \\
    \hline
    \end{tabular}
    \caption{Amplifier Consistency (1000 runs)}
    \label{tab:amp_consistency}
\end{table}

\subsection{Binary Search}
The binary search algorithm described in section \ref{binary_search} was implemented in various sizes using a similar testbench.  The testbench picks a random line in the set and runs the binary search algorithm to determine if it could accurately pick out the single present cacheline.  The results are shown in Table \ref{tab:binary_search}.

The binary search algorithm was very accurate for all sizes through 128 but did drop off in the 256 configuration.  With larger sizes the algorithm does run longer and this increases the risk of errors.

The performance of the binary search algorithm is slower than the simple approach of testing each cacheline sequentially by about 15-20x.  However, if the cost of measuring a cacheline is expensive (such as if amplification is required before the cacheline can be measured) then the binary search algorithm is a useful optimization as many fewer measurements are required.

\begin{table}[t]
    \centering
    \begin{tabular}{|c|c|c|}
    \hline
   
    Size & Correct Results & Accuracy \\
    \hline
    4 & 100000 & 100\% \\
    8 & 99999 & 100\% \\
    16 & 99995 & 100\% \\
    32 & 99999 & 100\% \\
    64 & 99991 & 99.99\% \\
    128 & 92374 & 92.37\% \\
    256 & 66398 & 66.40\% \\
    \hline 
        
    \end{tabular}
    \caption{Binary Search Results (100k runs)}
    \label{tab:binary_search}

\end{table}

\subsection{Cacheline Counter}
The cacheline counting algorithm described in section \ref{cacheline_counter} was also implemented in various sizes using a similar testbench.  The testbench randomly initialized the set of cachelines so some were and some were not present and then ran the cacheline counting algorithm to determine if it could correctly count the number of present cachelines.  For instance, for size 16 then a set of 16 cachelines were randomly initialized in various states and five cachelines was used to capture the 5-bit count result.  The results are shown in Table \ref{tab:cacheline_counter}.

The cacheline counter algorithm was highly accurate for the sizes tested, although as expected accuracy did decline slightly at larger sizes.

\begin{table}[t]
    \centering
    \begin{tabular}{|c|c|c|}
    \hline
   
    Size & Correct Results & Accuracy \\
    \hline
    4 & 99991 & 99.99\% \\
    8 & 99981 & 99.98\% \\
    16 & 99937 & 99.94\% \\
    32 & 99842 & 99.84\% \\
    64 & 99697 & 99.70\% \\
    128 & 99275 & 99.28\% \\
    256 & 97982 & 97.98\% \\
    \hline 
        
    \end{tabular}
    \caption{Cacheline Counter Results (100k runs)}
    \label{tab:cacheline_counter}

\end{table}

\section{Mitigations} \label{mitigations}
As this paper does not demonstrate a new attack, there is likely no need for mitigations specific to the gadgets discussed here, and such mitigations are unlikely to easily exist.  Many of the existing mitigations for side channel attacks like Spectre rely on preventing the undesired behavior in the first place\cite{msftspectre,currentspectremitigations,llvmslh}, including by preventing the attacker from mounting such an attack or isolating memory so it cannot be read even during speculation\cite{chromesiteisolation}.  These mitigations are effective and should remain as the primary defense.

The techniques demonstrated here do heavily rely on branch misprediction and as such, it may be possible to detect their use through performance monitor counters or other tools that can detect large amounts of mispredicted branches.  The challenge with this approach is that mispredicted branches are not uncommon and being able to reliably differentiate between attack code and benign code may be quite difficult.

As this paper demonstrates, reducing the precision of timers available to an attacker does not render cache-based side channel attacks impossible, but it does require extra work to employ techniques like signal amplification.  And if an attacker is required to use signal amplification, the performance of their attack may drop dramatically as each signal may take milliseconds to read instead of nanoseconds.  Therefore, reducing the precision of timers may still be a useful defense-in-depth strategy, but it does not eliminate the threat of these side channels attacks.

\section{Future Work}

Optimizing cache signal recovery is rather unexplored area, and future research is needed to determine the practical limits of signal modification and amplification in other contexts, including browsers, as well as evaluation on non-x86 microarchitectures.

Beyond side channel attacks, computation on the presence of cache lines could be an interesting new paradigm for confidential computing.  As secret data can be stored in the state of cachelines instead of in registers or memory, this information would be invisible to a higher privilege entity like an operation system.  Furthermore, attempts to conduct side channel attacks \textit{against} this type of software may prove futile as any modification of the cache state may corrupt the data being used by the program.  As such, a program would be aware if someone attempted to read its state, which is a potentially unique security property.

Finally, error detection and correction with cache side channel signals may be a useful area of exploration.  Although each operation on cache state carries with it some risk of inaccuracy, this may be able to be mitigated using error correction techniques.

\section{Conclusion}
This paper demonstrates a number of new speculation-based gadgets for optimizing and amplifying the signals from cache-based covert and side channel attacks.  Being able to modify signals without measuring them is a new enabling technology for advanced side channel research, especially in cases where signals are either difficult to measure due to a lack of high precision timer sources, or if access to the timer may be monitored by an adversary.  Algorithms such as the binary search and cacheline counter can be employed to reduce the number of timing measurements needed by the attacker, thus improving overall efficiency or avoiding detection.

For real-world scenarios involving attacks which cannot be easily repeated (such as ones requiring specific timing windows), the techniques shown in this paper are likely to be especially effective as signal modification and amplification can be achieved without having to invoke the victim multiple times.  A single signal can be amplified to the point of being measured reliably with timers as coarse as 100$ms$ or more, which are typically easily available.

Understanding these techniques is important for both attackers and defenders, and will hopefully lead to improved mitigations as well as additional research in the area of side channel signal recovery.

{ \normalsize \bibliographystyle{IEEEtran} 
\bibliography{main.bbl}}

\end{document}